# Momentum-Space-Engineered Spatial Photonic Ising Machine for Long-Range Interactions

Haijun Zhou[1], Hengyang Li[1], Maolin Wang[1, *], Yunru Chen[1], Yingxiong Qin[1], Xiahui Tang[1], Gang Xu[1]
[1]*School of Optical and Electronic Information, Huazhong University of Science and Technology, Wuhan, China.*
* maolin_w@hust.edu.cn

**Abstract**: Long-range Ising models (LRIMs) with dense nonlocal and competing interactions are central to statistical physics, quantum simulation, and complex networks. Although the spatial photonic Ising machine (SPIM) exploits intrinsic optical parallelism for Ising computation, its ability to faithfully encode dense long-range couplings and capture the resulting thermodynamic signatures remains underexplored. Here, we present a momentum-space-engineered SPIM framework that maps prescribed long-range coupling kernels onto momentum-space masks for parallel Hamiltonian evaluation. Based on a high-fidelity optical field propagation model, the annealing dynamics of LRIMs with power-law and Ruderman-Kittel-Kasuya-Yosida (RKKY) interactions are systematically investigated. For the power-law model, we investigate the modulation of the estimated critical temperature by the decay exponent $\sigma$ and coupling cutoff radius $R$. For the RKKY model, we reproduce diverse ordered states induced by complex competing long-range interactions. A proof-of-principle experiment demonstrates the physical feasibility of our approach. This framework broadens the class of many-body systems accessible to the SPIM platform.

## 1. Introduction

Long-range interactions constitute a ubiquitous form of nonlocal coupling in both natural and engineered systems. Unlike short-range couplings confined to local neighbors, long-range interactions sustain strong correlations between distant constituents, inducing unique physical phenomena absent in short-range systems, such as non-additivity, ensemble inequivalence, and complex spatially ordered structures driven by intense competition [1]. Consequently, they play a pivotal role in various complex systems, including fluid dynamics [2], quantum systems [3], and statistical physics [4].

Among various long-range interacting models, the long-range Ising models (LRIMs) with power-law [5] and Ruderman-Kittel-Kasuya-Yosida (RKKY) interactions [6] are two widely studied examples. In the power-law interaction, the decay exponent $\sigma$ dictates the non-locality of spin coupling, influencing the critical temperature, ground-state configurations, and non-equilibrium relaxation processes [7]. This interaction manifests across a wide spectrum of physical systems, including gravitational systems [8], charged plasmas [9], and dipolar magnets [10]. On the other hand, the RKKY indirect exchange interaction serves as a prototypical mechanism in magnetic systems, where the effective coupling between localized magnetic moments is mediated by conduction electrons [6]. Because its coupling strength oscillates and decays with distance, spin pairs at varying separations favor either ferromagnetic or antiferromagnetic alignment, thereby naturally introducing long-range competition. Widespread in dilute magnetic alloys [11], rare-earth metals [12], and magnetic semiconductors [13, 14], the RKKY interaction constitutes a vital mechanism for understanding complex magnetic order and frustrated structures. Consequently, engineering and manipulating these long-range interactions within a controllable platform is important for exploring the ground-state phase diagrams and phase transition behaviors of complex LRIMs.

However, since each spin in the LRIM simultaneously couples with a vast number of distant spins, the dense network of interactions causes the computational overhead of energy evaluation to scale quadratically with the system size $(O(\mathrm{N}^2))$. Consequently, classical electronic computers encounter severe bottlenecks in dynamical simulations and ground-state searches for large-scale, long-range spin systems. To circumvent this limitation, advanced algorithms have been developed [15], and long-range interacting systems have been extensively investigated across various physical platforms [16-18]. Among these, the spatial photonic Ising machine (SPIM) [19-23] has emerged as a promising optical computing paradigm, capable of efficiently simulating and solving large-scale LRIMs by exploiting the parallel propagation and interference of optical fields. To date, SPIMs have been deployed to investigate condensed-matter phenomena, such as phase transitions [24] and replica symmetry breaking [25]. Wavelength [26], spatial [27,28], and temporal multiplexing [29] have expanded SPIM programmability for tackling various NP-hard problems [30-35]. Recently, the deployment of frequency-domain spatial filtering has enabled the simulation of spin-glass and $J_1$-$J_2$-$J_3$ Ising models within SPIM platforms [36,37], thereby demonstrating the efficacy of this approach in programmable coupling design and the physical realization of Ising Hamiltonians. More recently, quadratic unconstrained binary optimization (QUBO) with spatially convolutional structures has been represented as a two-dimensional periodic spatial QUBO, which can be implemented on SPIMs without multiplexing [38]. Nevertheless, existing optical simulation paradigms predominantly focus on neighboring coupling topologies or QUBO problems with convolutional formulations. Utilizing the SPIM platform to physically realize dense long-range interactions featuring power-law and RKKY profiles and to investigate their thermodynamic behavior and low-energy states has not yet been systematically explored.

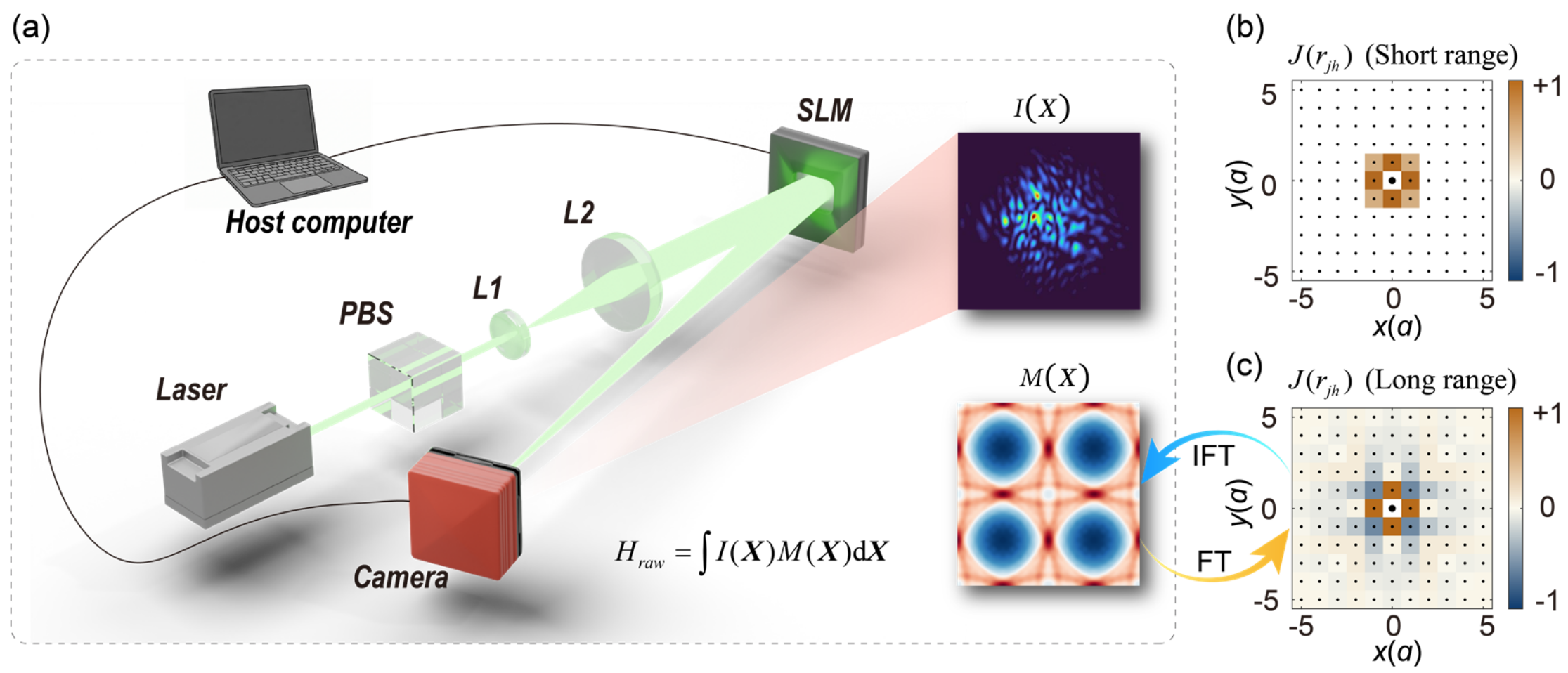

**Fig. 1.** (a) Schematic illustration of the system architecture of the SPIM designed for long-range interactions. $I(\boldsymbol{X})$ denotes the two-dimensional optical intensity distribution captured by the camera at the Fourier back focal plane. The spatially weighted integration of $I(\boldsymbol{X})$ with the momentum-space mask $M(\boldsymbol{X})$ enables the parallel evaluation of the Hamiltonian, while a host computer executes closed-loop algorithmic decisions and spin-phase updates. Additionally, a virtual Fresnel lens phase is superimposed onto the SLM to effectively eliminate the background noise of unmodulated zero-order light via the defocusing effect. (b) Short-range spin coupling matrix exhibiting localized characteristics. (c) Long-range spin coupling matrix exhibiting nonlocal characteristics, whose spatial configuration satisfies a Fourier transform relationship with the momentum-space mask.

In this research, we extend momentum-space engineering beyond such short-range interactions to dense LRIMs with power-law and RKKY coupling profiles, and further investigate their thermodynamic behaviors and low-energy states within the SPIM framework. Using an optical field-propagation model, we perform numerical simulations to track the dynamical annealing processes of long-range interacting systems. For the LRIM with power-law interaction, we elucidate how the decay exponent $\sigma$ and the coupling cutoff radius $R$ modulate the critical temperature. Crucially, for the RKKY-type LRIM, our framework reproduces distinct long-period stripe, stripe glass, antiferromagnetic, and super-antiferromagnetic phases at four representative Fermi wavevectors ($k_F$ = 0.2, 1.0, 2.2, and 3.4 $a^{-1}$), exhibiting excellent agreement with theoretical benchmarks. Finally, we construct a hardware SPIM platform and perform proof-of-principle experiments to validate the proposed momentum-space-engineered architecture for long-range coupling and demonstrate that the system can successfully reach the ground-state configurations of RKKY-type LRIM. To the best of our knowledge, this is the first experimental demonstration of ground-state configurations in RKKY Ising model on the SPIM platform. This work provides a versatile physical approach to investigate dense and frustrated long-range Ising systems.

## 2. Principles and Methods

Initially, without an external magnetic field, the Hamiltonian of the LRIM can be formulated as follows:

$$H = -\frac{1}{2}\sum_{j,h=1;j\neq h}^{N} J(r_{jh})s_j s_h, \qquad (1)$$

where $s_j = \pm 1$ represents the value of the $j$-th spin, $N$ is the total number of spins, and $J(r_{jh})$ denotes the interaction strength between spins $s_j$ and $s_h$. The distance between spins $s_j$ and $s_h$ is given by $r_{jh} = |\boldsymbol{r}_{jh}| = |\boldsymbol{x}_j - \boldsymbol{x}_h|$, where $\boldsymbol{x}_j = (x_j, y_j)$ represents the position vector of spin $s_j$. In this work, we focus on two prototypical categories of long-range interactions. The first category is the power-law interaction, whose coupling coefficient is expressed as $J(r_{jh}) = {r_{jh}}^{-(\sigma+d)}$, where $d$ denotes the spatial dimension and $\sigma$ is the decay exponent that governs the interaction range [40]. The second category is the RKKY indirect exchange interaction characterized by spatially alternating signs, with its spatial coupling coefficient defined as $J(r_{jh}) = \cos(2k_F\, r_{jh})/r_{jh}^d$ , where the Fermi wavevector $k_F$ dictates the spatial oscillation period of the coupling function. Tuning $k_F$ can modulate both the sign and relative strength of the coupling between spin pairs at varying separations, thereby inducing rich, competing frustrated ground-state configurations [41]. Fig. 1(b) and 1(c) compare the typical coupling matrices of short-range and long-range interactions, which visually demonstrate the nonlocal spatial correlations between a single spin and distant spin clusters within the LRIM framework.

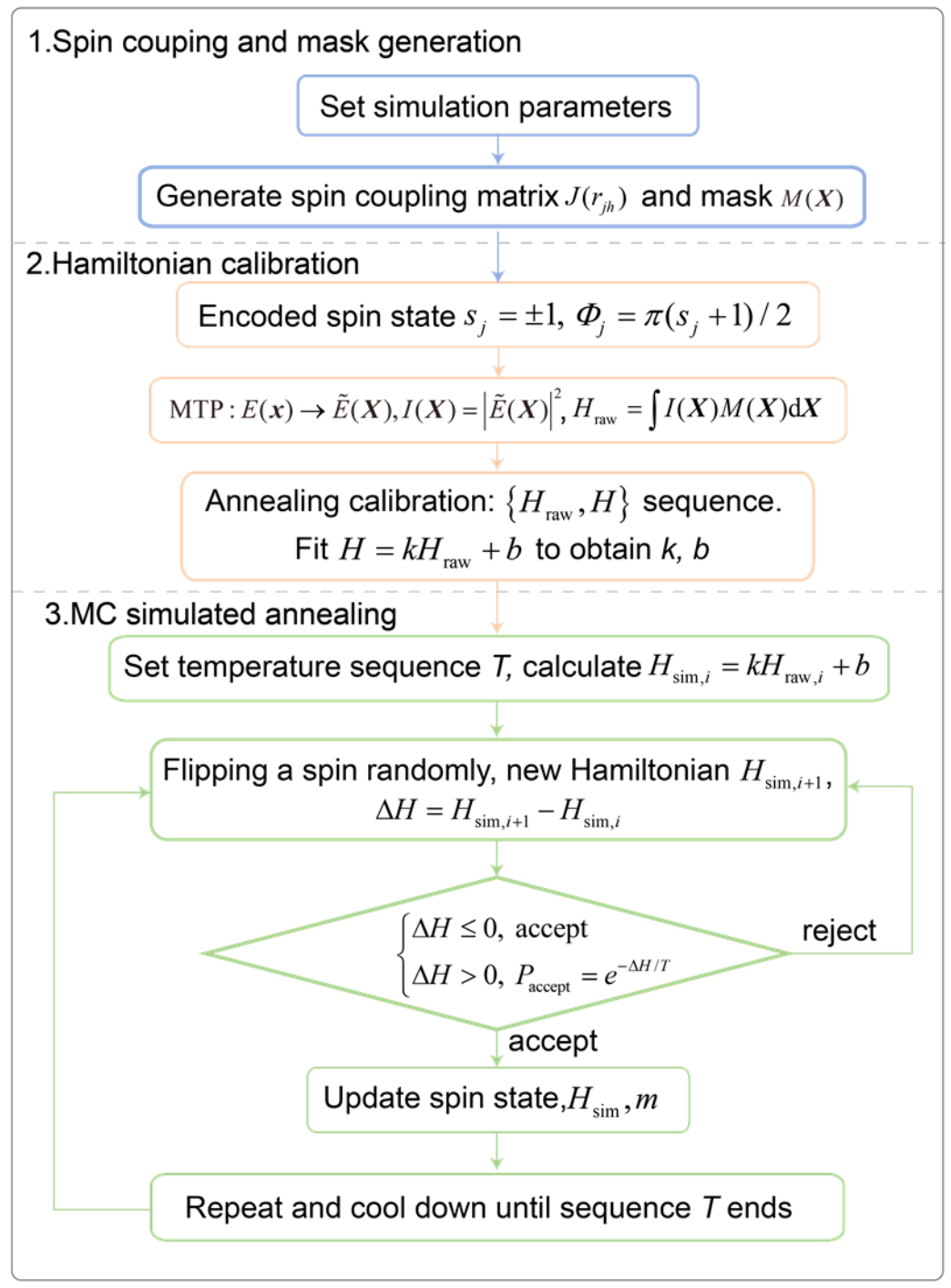

**Fig. 2.** Implementation workflow for the numerical simulation of optical field propagation in LRIMs. The procedure comprises three primary stages: (1) spin coupling configuration and momentum-space mask generation; (2) Hamiltonian calibration based on the MTP optical propagation model, which relates the raw simulated energy $H_{\mathrm{raw}}$ to the target Hamiltonian $H$ and compensates for finite-grid sampling errors; and (3) Metropolis MC simulated annealing for tracking the evolution of spin configurations toward low-energy states. $P_{\mathrm{accept}}$ denotes the acceptance probability.

The architecture and operating principle of the SPIM for implementing long-range interactions are illustrated in Fig. 1(a). In this system, the Ising spin configuration is encoded into the phase profile of the incident optical field by a spatial light modulator (SLM). Through parallel optical interference and diffractive propagation, the optical field at the Fourier plane provides an optical mapping and evaluation of the Ising Hamiltonian. Each individual spin is represented by a macro-pixel consisting of multiple SLM pixels, where the phase $\phi_j$ of the $j$-th macro-pixel takes a value of either 0 or $\pi$ to satisfy the phase modulation condition $e^{\mathrm{i}\phi_j} = s_j$, yielding a modulated complex field directly behind the SLM plane defined as $E(\boldsymbol{x}) = \sum_{j=1}^{N} A_j s_j \mathrm{rect}\big((\boldsymbol{x}-\boldsymbol{x}_j)/a\big)$ . Here, $\boldsymbol{x} = (x,y)$ denotes the spatial coordinate on the SLM plane, $A_j$ represents the amplitude, $a$ is the macro-pixel width that corresponds to the spin lattice constant, and $\mathrm{rect}(\boldsymbol{x}/a)$ denotes the aperture function of the pixel. Upon propagation through the lens, the optical field from the SLM plane undergoes a spatial Fourier transform at the back focal plane. Neglecting the slow spatial variations of the macro-pixel diffraction envelope and assuming an approximately uniform incident amplitude, the optical intensity distribution captured at the Fourier plane is expressed as $I(\boldsymbol{X}) = C_0 \sum_{j,h=1}^{N} s_j s_h \exp\left(-\mathrm{i}\frac{2\pi}{\lambda f}(\boldsymbol{x}_j - \boldsymbol{x}_h)\cdot \boldsymbol{X}\right)$, where $\boldsymbol{X} = (X, Y)$ denotes the spatial coordinate at the Fourier plane, $f$ is the focal length of the lens, $\lambda$ is the optical wavelength, and $C_0$ is a constant factor. By introducing a mask $M(\boldsymbol{X})$ at the Fourier plane and performing a weighted integration with the optical intensity $I(\boldsymbol{X})$, the raw optical energy is obtained as:

$$H_{\mathrm{raw}} = \int I(\boldsymbol{X}) M(\boldsymbol{X}) d\boldsymbol{X} = C_0 \sum_{j,h=1;j\neq h}^{N} J(\boldsymbol{r}_{jh}) s_j s_h + C_1, \tag{2}$$

where $C_1$ is a constant bias independent of the spin configuration, introduced by the self-interaction term $j = h$. The effective coupling kernel generated by the mask is given by $J(\boldsymbol{r}_{jh}) = \int M(\boldsymbol{X}) \exp(-\mathrm{i}\frac{2\pi}{\lambda f}\boldsymbol{r}_{jh}\cdot \boldsymbol{X}) \mathrm{d}\boldsymbol{X}$. Given that the long-range coupling function $J(\boldsymbol{r}_{jh})$ is real-valued and exhibits spatial inversion symmetry, satisfying $J(\boldsymbol{r}_{jh}) = J(-\boldsymbol{r}_{jh})$, we discretize it as a coupling matrix for numerical and optical implementation. Consequently, the corresponding momentum-space mask $M(\boldsymbol{X})$ can be further formulated as:

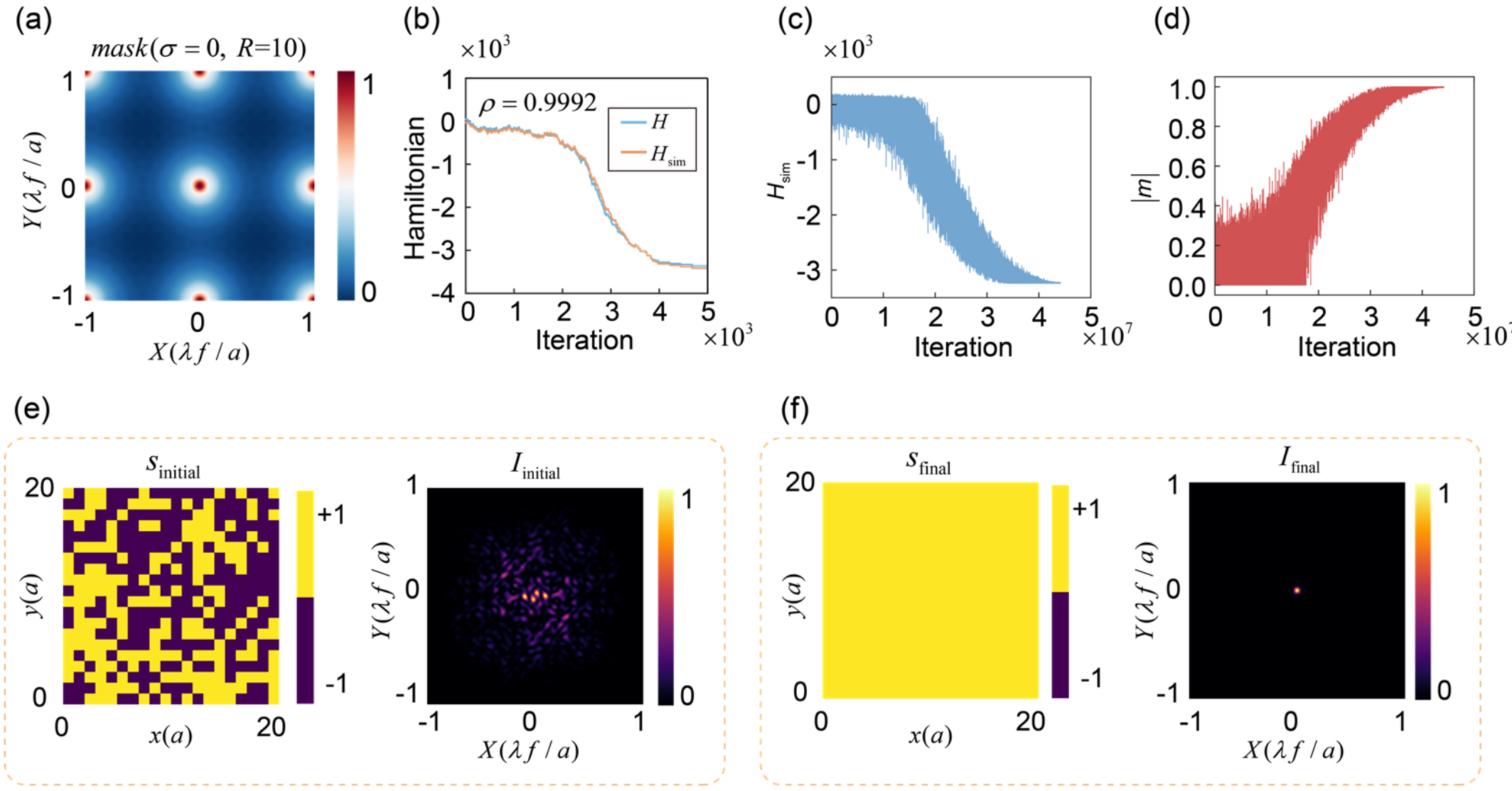

**Fig. 3.** Numerical simulation results for the power-law LRIM with $\sigma = 0$, $R = 10$. (a) Momentum-space mask distribution corresponding to the target coupling matrix. (b) Evolution curves of the simulated energy $H_{\text{sim}}$ and the target Hamiltonian $H$ during the Hamiltonian calibration stage, yielding a Pearson correlation coefficient of $\rho = 0.9992$. (c) and (d) Simulated Hamiltonian trajectory and absolute magnetization $|m|$ obtained during the subsequent annealing process. (e) Initial random spin configuration and the corresponding Fourier-plane intensity distribution. (f) Final ferromagnetic spin configuration after annealing convergence and the corresponding central focused diffraction spot at the Fourier plane.

$$M(\boldsymbol{X}) \propto \sum_{j<h}^{N} J(\boldsymbol{r}_{jh}) \cos\left(\frac{2\pi}{\lambda f}\boldsymbol{r}_{jh} \cdot \boldsymbol{X}\right). \qquad (3)$$

Consequently, given the coupling matrix and the interaction range of the long-range interactions, the corresponding momentum-space mask $M(\boldsymbol{X})$ can be constructed according to Eq. (3). Subsequently, the mask is normalized via $[M(\boldsymbol{X}) - M_{\text{min}}]/(M_{\text{max}} - M_{\text{min}})$, with the resulting scaling effects compensated for during the subsequent Hamiltonian correction stage. This mechanism directly maps complex, nonlocal spin interactions from real space onto the spatial function distribution of the momentum-space mask. During practical energy evaluation, by executing a spatially weighted integration of the captured two-dimensional Fourier-plane optical intensity with the mask, the computing architecture achieves single-shot parallel evaluation of the long-range Ising Hamiltonian directly at the physical optical level.

## 3. Numerical Simulations

Leveraging the aforementioned theoretical framework, we conduct numerical simulations of optical field propagation to investigate the dynamical annealing processes of the LRIMs with power-law and RKKY interactions, respectively, on a two-dimensional (2D) square lattice ($d = 2$). The overall simulation workflow is illustrated in Fig. 2. To achieve high-fidelity simulation of optical field propagation and discrete sampling within the SPIM, we implement the matrix triple product (MTP) algorithm [39], which completely decouples spatial and frequency-domain computational grids to match arbitrary camera pixels and target mask geometries. Owing to its inherently matrix-based formulation, the MTP framework is highly amenable to GPU parallel acceleration. Furthermore, leveraging the linearity of the MTP enables a localized incremental update scheme during single-spin flips, significantly accelerating the computational rate by eliminating redundant full-field transformations.

Based on the aforementioned simulation workflow, we first investigate the LRIM with the power-law interaction $J(r_{jh}) = {r_{jh}}^{-(\sigma+2)}$. The specific parameters utilized in the numerical simulations are as follows: the SLM pixel width $L_{\text{SLM}} = 8\ \mu\text{m}$, the camera pixel width $L_{\text{cam}} = 3.76\ \mu\text{m}$, the laser wavelength $\lambda = 532$ nm, and the focal length of the lens $f = 1000$ mm. The total number of spins in the system is $N = 400$, with each spin consisting of $50 \times 50$ SLM pixels. The representative simulation profiles under a decay exponent of $\sigma = 0$ and a coupling cutoff radius of $R = 10$ ($J(r_{jh}) = {r_{jh}}^{-(\sigma+2)}$ for $0 < r_{jh} \le R$ and $J(r_{jh}) = 0$ for $r_{jh} > R$) are presented in Fig. 3. The Pearson correlation coefficient between the simulated energy $H_{\mathrm{sim}}$ and the target Hamiltonian $H$ yields $\rho = 0.9992$, demonstrating that $H_{\mathrm{sim}}$ acts as a highly accurate linear proxy for $H$. The Monte Carlo (MC) simulated annealing algorithm is then used to track the dynamical evolution of the system from a high-temperature disordered state to a low-temperature ordered state. As shown in Fig. 3(e), the spin configuration is randomly distributed initially, resulting in a broadly dispersed Fourier-plane intensity profile. As the simulated annealing process progresses, the system gradually evolves toward a ferromagnetic ordered state; accordingly, the corresponding Fourier-plane intensity distribution rapidly contracts toward the center, ultimately focusing into a sharp diffraction spot. The final absolute magnetization of the system reaches $|m| = \left|\sum_{j=1}^{N} s_j\right|/N$=1, indicating convergence to the ferromagnetic state.

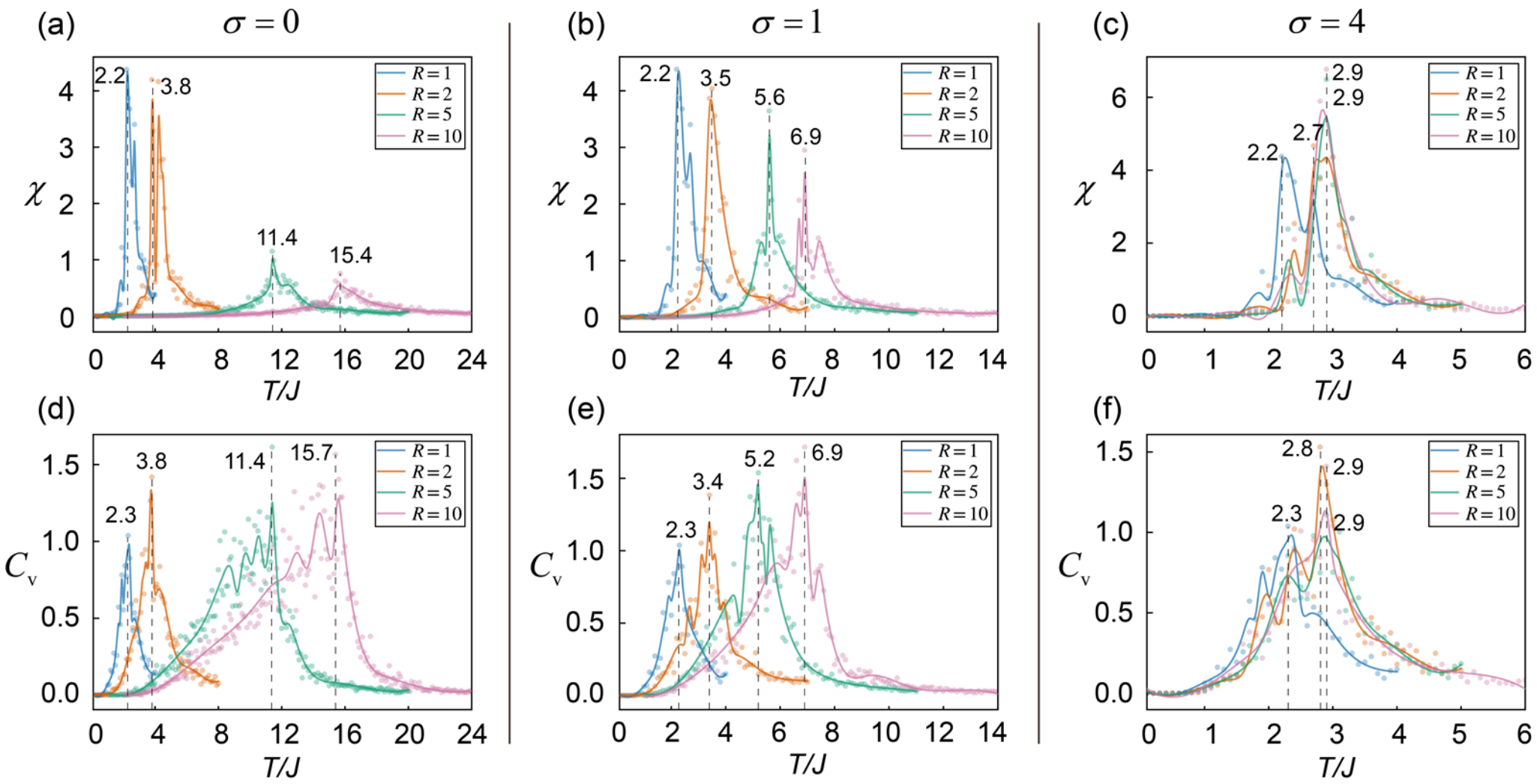

**Fig. 4.** Numerical simulation results of the LRIM with a power-law interaction $J(r_{jh}) = r_{jh}^{-(\sigma+2)}$. (a)–(c) Evolution of the magnetic susceptibility $\chi$ and (d)–(f) specific heat capacity $C_v$ as functions of the normalized temperature $T/J$ under various combinations of the decay exponent $\sigma$ ($\sigma = 0, 1, 4$) and the coupling cutoff radius $R$ ($R = 1, 2, 5, 10$). The discrete markers represent the raw Monte Carlo simulation data, while the solid lines denote the corresponding smoothed fitting curves (obtained via cubic smoothing splines). The vertical dashed lines and labeled numerical values indicate the estimated critical temperatures, which are determined directly by the maximum positions of the raw discrete simulation data. The simulations are performed on a 2D square lattice with a normalized temperature step size of $\Delta(T/J) = 0.1$. At each temperature point, sampling is executed after 100,000 thermalization spin-flip attempts, followed by a sampling interval of 5 MC sweeps.

We further investigate the effects of $\sigma$ and $R$ on the phase-transition behavior of the system. At each temperature step $T$, the magnetization $m$, magnetic susceptibility $\chi = N[\langle m^2\rangle_T - \langle m\rangle_T^2]/(k_B T)$, and specific heat capacity $C_v = [\langle H_{\mathrm{sim}}^2\rangle_T - \langle H_{\mathrm{sim}}\rangle_T^2]/(N k_B T^2)$ are quantitatively evaluated from the sampled spin configurations $\{s_j\}$. Here, the Boltzmann constant is set to $k_B = 1$, and $\langle\cdot\rangle_T$ denotes the statistical ensemble average over the spin states captured at temperature $T$. The peak positions of $\chi$ and $C_v$ are used to estimate the temperature of the phase transition. To improve equilibration and reduce correlations between sampled spin configurations, sampling is performed only after sufficient thermalization, with an interval of 5 MC sweeps ($5N$ spin-flip attempts) between successive samples. Fig. 4 displays the scatter plots and corresponding fitting curves of $\chi$ and $C_v$ under various combinations of $\sigma$ and $R$. At $R = 1$, the system degenerates into the nearest-neighbor model, where the transition temperatures determined by $\chi$ and $C_v$ are $2.3J$ and $2.2J$, respectively. These values are in excellent agreement with the exact Onsager solution ($T_c \approx 2.269J$), thereby validating the accuracy of the simulation framework. As $R$ increases, the characteristic peaks overall shift toward higher temperatures, indicating that long-range ferromagnetic coupling enhances the ordering tendency of the system, with this effect being most pronounced in the zero-decay limit ($\sigma = 0$). Conversely, as $\sigma$ increases, the long-range coupling decays rapidly. At $\sigma = 4$, the peak positions for various $R$ ($R = 2, 5, 10$) lie within $2.7J$–$2.9J$, showing a weaker dependence on $R$ over the tested range. The reduced sensitivity to $R$ suggests that, within the finite system size and parameter range considered, rapidly decaying nonlocal couplings make a progressively smaller contribution to the thermodynamic behavior. Consequently, the system exhibits an increasing tendency toward short-range-dominated behavior.

To extend the momentum-space mask method to RKKY-type LRIM featuring spatial oscillations, we first employ numerical calculations at a purely theoretical level to establish a standard baseline for the system, thereby providing a clear theoretical reference for numerical simulations of optical field propagation and delineating the panoramic converged-state landscape. Scanning the normalized Fermi wavevector $k_F$ from 0 to 3.5 in steps of 0.1 (in units of $a^{-1}$) yields the converged -state phase diagram of the RKKY system illustrated in Fig. 5. This benchmark phase diagram demonstrates that the continuous modulation of the spatial proportion between long-range ferromagnetic and antiferromagnetic interactions drives the system to evolve exotic ordered structures characterized by multi-phase competition, which stems from intense nonlocal competition and interaction-induced frustration.

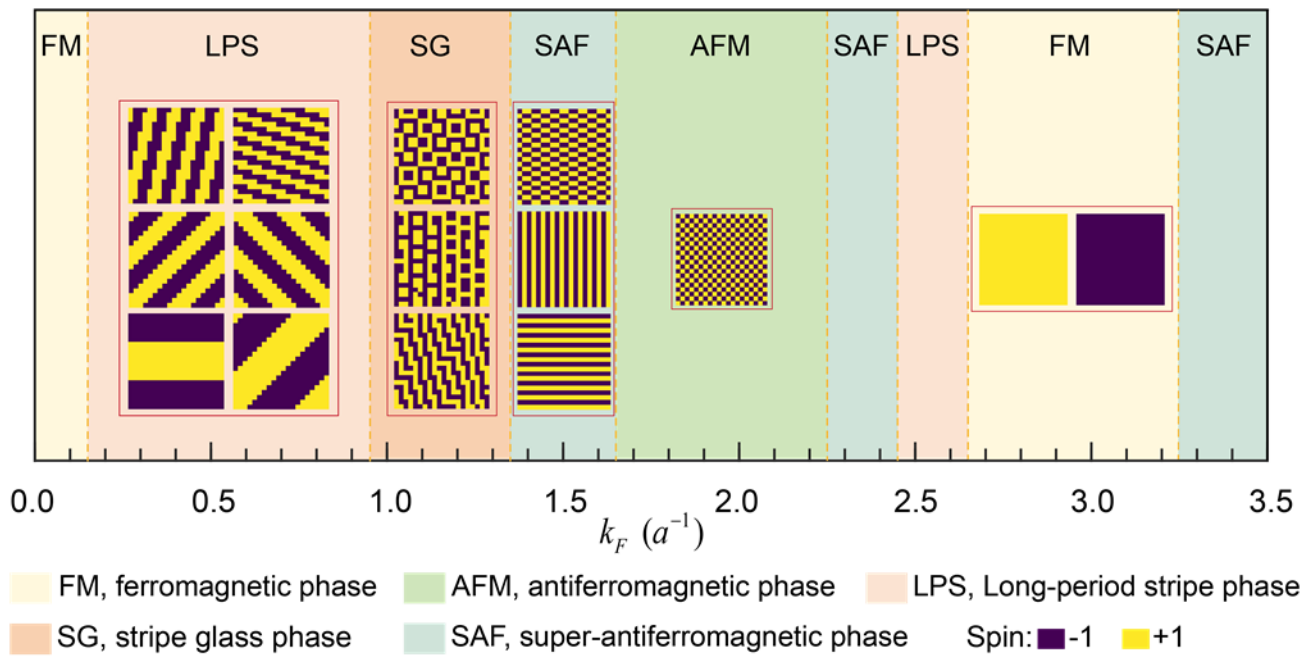

**Fig. 5.** Theoretical converged-state phase diagram of the LRIM with RKKY interaction $J(r_{jh}) = \cos(2k_F r_{jh})/r_{jh}^2$. The diagram, obtained via numerical simulations based on theoretical formulations, maps five distinct nonlocal spin-ordered phases, where the insets display the representative spin configurations within each regime. These phases encompass the ferromagnetic phase (FM, $k_F \in [0, 0.1] \cup [2.7, 3.2]\ a^{-1}$), the long-period stripe phase (LPS, $k_F \in [0.2, 0.9] \cup [2.5, 2.6]\ a^{-1}$) featuring spatially periodic spin domains modulated by competing long-range interactions, and the stripe glass phase (SG, $k_F \in [1.0, 1.3]\ a^{-1}$) characterized by the competitive coexistence of multiple local stripe arrangements. The remaining regimes correspond to the super-antiferromagnetic phase (SAF, $k_F \in [1.4, 1.6] \cup [2.3, 2.4] \cup [3.3, 3.5]\ a^{-1}$) and the antiferromagnetic phase (AFM, $k_F \in [1.7, 2.2]\ a^{-1}$).

Subsequently, to validate the applicability and accuracy of the proposed method for LRIM with RKKY interaction $J(r_{jh}) = \cos(2k_F r_{jh})/r_{jh}^2$, we select four representative wavevector parameters ($k_F$ = 0.2, 1.0, 2.2, and 3.4 $a^{-1}$) from different characteristic phase regimes of the aforementioned benchmark phase diagram to perform numerical simulations. For each parameter, the previously described Hamiltonian calibration and MC simulated annealing procedures are employed to obtain the corresponding low-energy spin configuration. Fig. 6 illustrates the coupling functions, the corresponding normalized masks, converged-state spin configurations, and spin structure factors under these four typical cases. The spin structure factor is defined as $S(\boldsymbol{k}) = \left|\sum_{j=1}^{N} s_j \exp(-\mathrm{i}\boldsymbol{k} \cdot \boldsymbol{x}_j)\right|^2/N$, where $\boldsymbol{k}$ denotes the reciprocal-space wavevector that satisfies the specific relationship $\boldsymbol{k} = 2\pi\boldsymbol{X}/(\lambda f)$. At $k_F = 0.2$, the effective coupling is dominated by positive ferromagnetic interactions at short distances and decays slowly with distance, driving the system to evolve into a long-period stripe phase with wide domains [41,42], whose reciprocal-space spin structure factor exhibits distinct characteristic double peaks in the low-wavevector region. As $k_F$ shifts to 1.0, the coupling exhibits a pronounced spatial sign-alternating characteristic, where intense nonlocal competition drives the system into the stripe glass phase [43]. Consequently, a local stripe structure lacking a single dominant period emerges in real space, and the structure factor displays multiple discrete peaks, reflecting the competitive coexistence of various nearly degenerate modes. When $k_F$ reaches 2.2, the rapidly oscillating effective coupling drives adjacent spins into an alternating antiparallel alignment, enabling the system to successfully converge to a standard antiferromagnetic ground state. The corresponding structure factor exhibits prominent peaks located at the reciprocal-space lattice points of $(\pm\pi/a,\ \pm\pi/a)$. Finally, at $k_F = 3.4$, the oscillation period of the coupling function shortens further, and the competition between ferromagnetic and antiferromagnetic coupling drives the system into a super-antiferromagnetic phase, where the structure factor exhibits a sharp characteristic peak at $(\pm\pi/a,$ $0)$.

## 4. Experiments and Results

In order to validate the aforementioned theoretical predictions and numerical simulations, we constructed a proof-of-principle experimental setup of the SPIM, as illustrated in Fig. 1(a). A continuous-wave laser operating at 532 nm served as the light source. The beam passed through a polarizing beam splitter (PBS) and a beam expander ($f_1$ = 15 mm and $f_2$ = 400 mm lens pair) to achieve a uniformly illuminated spot of approximately 40 mm in diameter on a reflective SLM (Santec SLM-210, 1920 × 1200 pixels, pixel width $L_{\mathrm{SLM}}$ = 8 μm). The SLM executed binary phase encoding of the Ising spin configuration, and the Fourier-plane intensity distribution was captured by a CMOS camera (ToupTek ATR533M, 3008 × 3008 pixels, pixel width $L_{\mathrm{cam}}$ = 3.76 μm). To suppress the undesired unmodulated light, a virtual Fresnel lens phase ($f$ = 1000 mm) was loaded onto the SLM so that the unmodulated light remained defocused and fell below the camera dark-field level under the experimental exposure conditions.

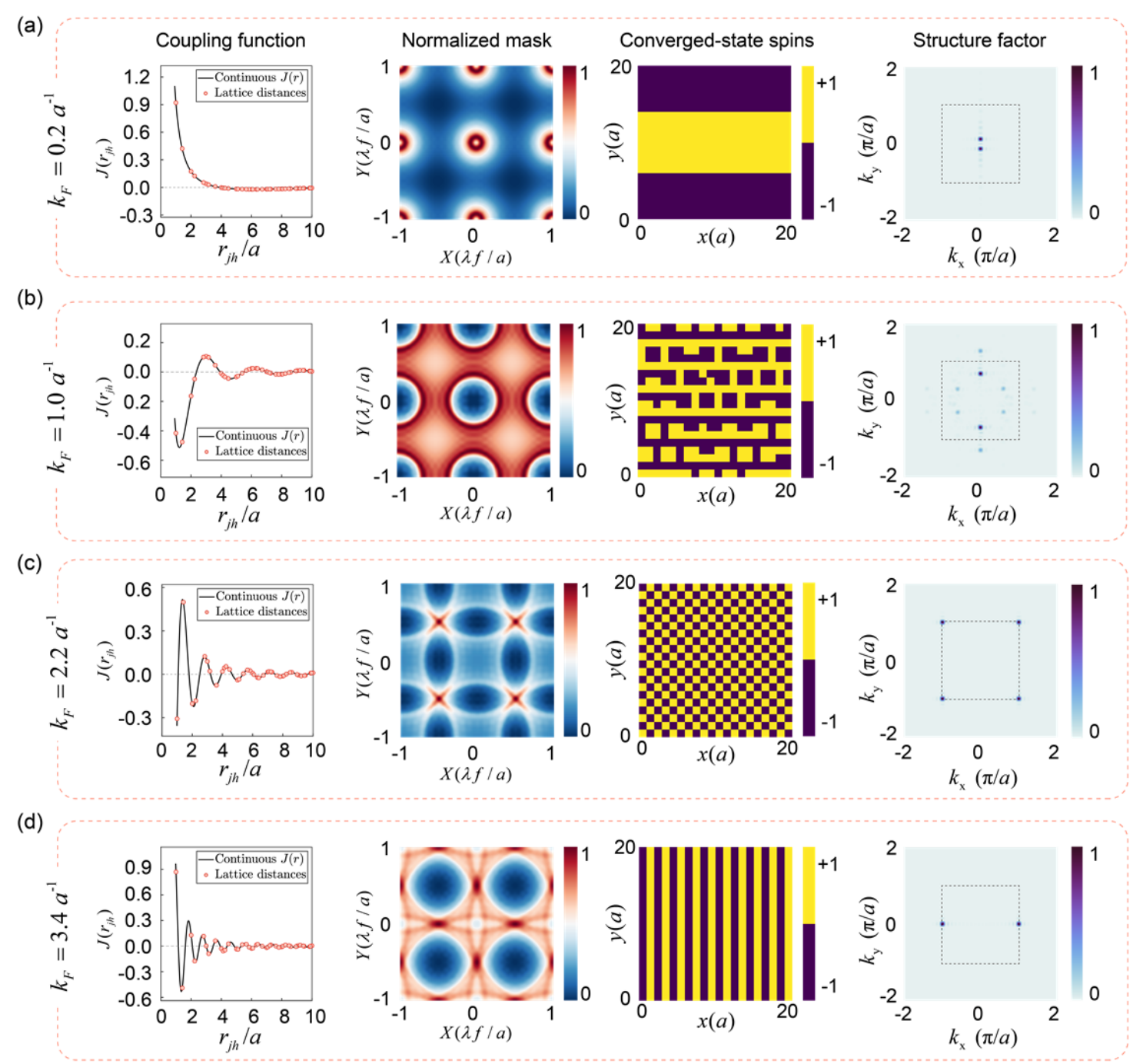

**Fig. 6.** System configurations and numerical simulation results of the LRIM framework with RKKY-type interaction $J(r_{jh}) = \cos(2k_F r_{jh})/r_{jh}^2$ across distinct interaction regimes. (a)–(d) Characteristic converged-state behaviors under four typical values of the Fermi wavevector ($k_F$ = 0.2, 1.0, 2.2, and 3.4 $a^{-1}$, respectively). From left to right, each row displays the radial dependence of the spin coupling function $J(r_{jh})$ on the normalized distance $r_{jh}/a$, the distribution of the normalized momentum-space mask, the converged-state spin configuration, and the corresponding distribution of the spin structure factor $S(\boldsymbol{k})$ in reciprocal space. The black dashed squares in the structure factor plots delineate the boundaries of the first Brillouin zone.

Using the aforementioned hardware platform, we carried out experimental validation for the RKKY-type long-range interactions. With a spin lattice scale of $N = 20 \times 20$, each spin was encoded by a $50 \times 50$-pixel macro-pixel with a physical width of $a = 50 \times L_{\mathrm{SLM}}$, and the resulting binary phase pattern was loaded onto the SLM. To align the experimentally captured Fourier-plane intensity region with the theoretically designed momentum-space mask $M(\boldsymbol{X})$, spatial calibration of the Fourier plane was performed beforehand. First, a two-dimensional checkerboard binary phase grating with a period of $a$ [Fig. 7(a)] was loaded onto the SLM. It generates four first-order diffraction spots [Fig. 7(b)] at the back focal plane, with theoretical Fourier-plane spatial coordinate $(\pm\lambda f/a, \pm\lambda f/a)$. Ideally, the four spot centers form a square with equal horizontal and vertical separations. In practice, slight rotation and scaling distortions may arise from optical misalignment and imaging aberrations. We therefore carefully optimized the optical alignment to minimize these distortions. The remaining deviations, arising from the limited precision of mechanical adjustment, were then compensated for using a two-dimensional affine transformation from the camera-pixel coordinates to the theoretical momentum-space coordinates. This calibration accurately maps the measured intensity distribution onto the coordinate grid defined by the designed mask, ensuring precise spatial alignment for subsequent weighted integration and Hamiltonian evaluation. To compensate for potential slow drifts in spot positions caused by environmental temperature fluctuations and mechanical stress relaxation during the experiment, the automation program re-identified the four first-order diffraction spot centers every 1000 iterations, updating the Fourier-space sampling grid in real time.

In the experiment, the momentum-space mask was implemented digitally in the electronic domain. After each camera exposure, the recorded two-dimensional Fourier-plane intensity was multiplied element-wise by the predefined digital momentum-space mask within the calibrated sampling region and then summed to obtain the experimental optical energy $H_{\mathrm{exp}}$. The digital implementation provides high spatial resolution and stable operation, while the corresponding element-wise multiplication and summation introduce only a small computational overhead. Based on the above method and the SPIM platform, we carried out experiments on RKKY-type LRIMs under different Fermi wavevectors $k_F$. Among the tested conditions, $k_F = 3.4a^{-1}$ exhibits the most pronounced spatial oscillations of the RKKY interaction and the strongest competition between ferromagnetic and antiferromagnetic interactions. We therefore selected this case for representative experimental demonstration (Additional experimental data obtained under different $k_F$ values, together with the corresponding visualized results, are provided in the research data repository [44]). Under this highly competing regime, the annealing calibration was executed to establish a linear mapping between the experimental optical energy $H_{\text{exp}}$ and the target Hamiltonian $H$. As illustrated in Fig. 7(c), the correlation coefficient reaches $\rho = 0.9942$, demonstrating that $H_{\text{exp}}$ faithfully characterizes $H$.

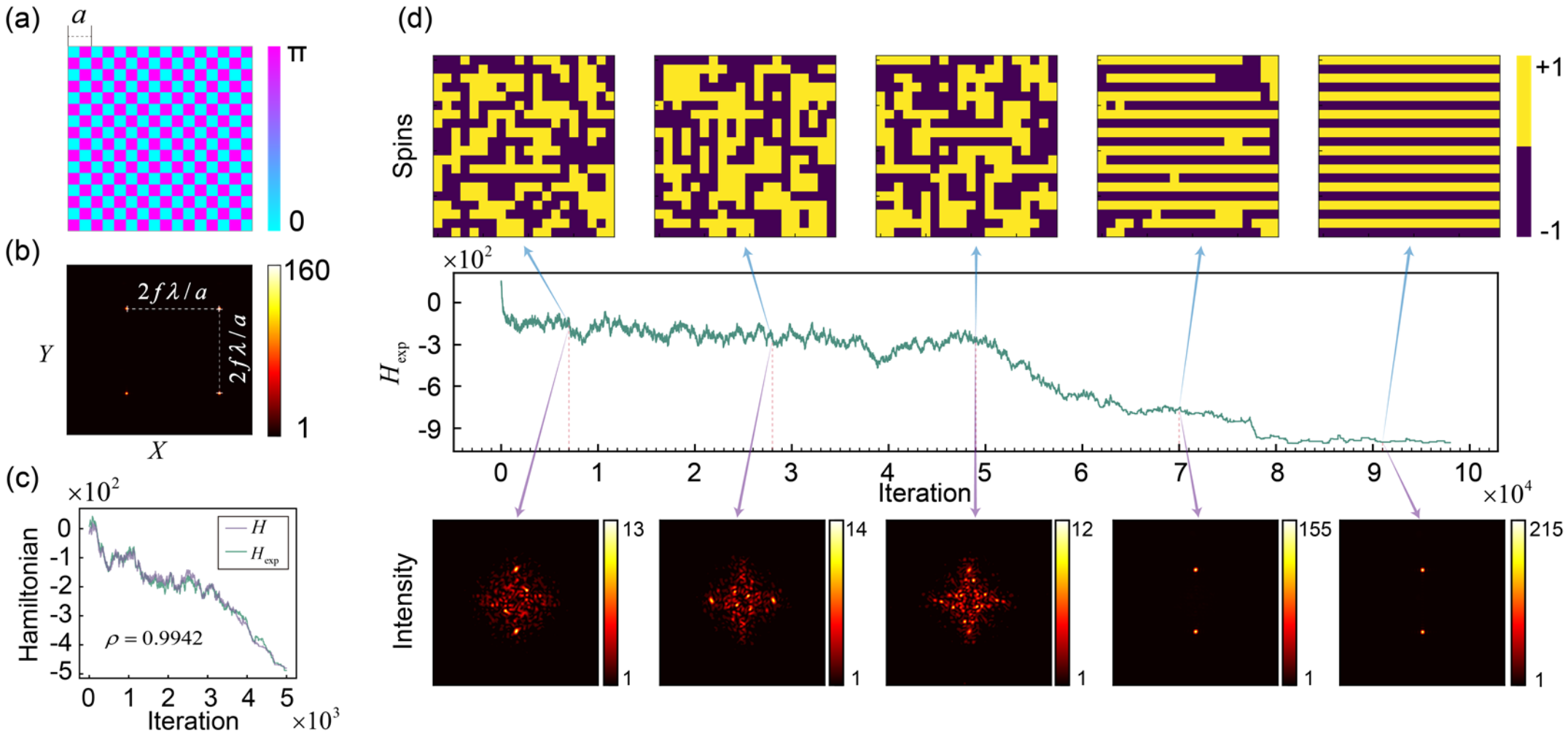

**Fig. 7.** Experimental verification results of the RKKY-type LRIM at $k_F = 3.4\ a^{-1}$ on the SPIM platform. (a) and (b) Spatial calibration of the Fourier plane, displaying the two-dimensional checkerboard binary phase pattern loaded onto the SLM and the corresponding first-order diffraction spots captured on the camera plane for precise sampling region alignment. (c) Hamiltonian scaling curves showing the linear tracking between $H_{\text{exp}}$ and $H$ with a Pearson correlation coefficient of $\rho = 0.9942$. (d) Dynamic experimental annealing trajectory, displaying the energy evolution curve alongside the transient spin configurations and corresponding Fourier-plane intensity distributions captured at 7,000, 28,000, 49,000, 70,000, and 91,000 iterations.

For the experimental annealing process, we employed a Metropolis simulated annealing scheme with an initial temperature of $T = 8.0J$ and a linear cooling schedule down to $T = 0.001J$, comprising 14 temperature steps in total. At each temperature step, 7000 single-spin-flip attempts were performed. For each attempt, one spin was randomly selected and flipped, and the corresponding Fourier-plane intensity distribution was recorded by the camera. The captured intensity was then multiplied element-wise by the predefined momentum-space mask and spatially summed to obtain the experimental optical energy $H_{\text{exp},i+1}$. The energy change $\Delta H = H_{\text{exp},i+1} - H_{\text{exp},i}$ was then calculated, and the Metropolis criterion was used to determine whether the spin flip was accepted. At the onset of annealing, the initial spin configuration was random, yielding a highly dispersed Fourier-plane intensity profile. As the annealing schedule progressed, the experimental optical energy decreased overall amidst local fluctuations, and a direction-selective stripe structure gradually emerged. Ultimately, the spin configuration converged to a well-ordered super-antiferromagnetic phase, where the Fourier-plane optical intensity evolved from the initial random speckle pattern into concentrated bright spots at the corresponding spatial frequencies. The overall evolution and the final converged state are consistent with the corresponding numerical simulation, and similar agreement was observed for the experimental results obtained at other Fermi wavevectors. Notably, minor baseline fluctuations persisted in the experimental optical energy during the final stabilization phase, primarily due to typical environmental perturbations such as transient laser power drift and ambient micro-vibrations. These residual fluctuations did not lead to a deviation of the spin configuration from the corresponding theoretical prediction during the final stabilization stage. Overall, the experimental results validate the feasibility of the proposed momentum-space-engineering approach for realizing and evaluating long-range Ising Hamiltonians on the SPIM platform.

## 5. Conclusion

In conclusion, we have proposed and experimentally verified a momentum-space-engineered SPIM framework for evaluating LRIM Hamiltonians and simulating LRIM annealing. By mapping complex nonlocal coupling onto a single momentum-space mask, this architecture enables highly parallel, single-shot Hamiltonian evaluations. Through numerical simulations of optical field propagation, our framework successfully reproduces the critical crossover in power-law interaction and reveals a rich converged-state phase diagram in the RKKY system featuring five distinct nonlocal spin-ordered phases driven by complex competing interactions. Furthermore, we experimentally verify the feasibility of applying the momentum-space masking method to long-range interaction models, validating the optical energy evaluation architecture on a proof-of-principle hardware platform.

Although the proposed momentum-space-engineering approach provides a direct and efficient means of implementing translationally invariant long-range interactions in SPIMs, extending its coupling programmability to more general interactions requires a more sophisticated Sidon-set-based spatial encoding strategy [45]. By assigning unique spatial displacements to different spin pairs, this encoding strategy allows distinct coupling strengths to be programmed for individual spin pairs, providing a potential route toward high-resolution pairwise coupling programmability. Meanwhile, further advances in optical and optoelectronic technologies could improve the operating speed and scalability of the proposed architecture. Emerging thin-film lithium-niobate electro-optic spatial light modulators [46], which have demonstrated MHz-rate programmable spatial modulation at the device level, provide a promising route toward rapid spin encoding. Combined with physical optical masks and high-dynamic-range point detectors, this architecture could reduce the bottlenecks imposed by SLM refresh rates and camera image transmission, thereby substantially accelerating the closed-loop iteration rate. Consequently, it provides a promising paradigm for the optical simulation and optimization of densely coupled, long-range frustrated Ising systems.

## Declaration of competing interest

The authors declare that they have no known competing financial interests or personal relationships that could have appeared to influence the work reported in this paper.

## Data availability

The underlying data and code supporting the findings of this study are openly available in the Zenodo repository [43].

## Acknowledgments

This work is partially supported by the Technique Innovation Key Project of Hubei Province in China (2025BAA004) and the National Natural Foundation of China (62275097).